# The significance of fuzzy boundaries of the barrier regions in single-molecule measurements of failed barrier crossing attempts.

Alexander M. Berezhkovskii[1] and Dmitrii E. Makarov[2]

[1]*Section of Molecular Transport, Eunice Kennedy Shriver National Institute of Child health and Human Development, National Institutes of Health, Bethesda, Maryland 20819, USA*

[2]*Department of Chemistry and Oden Institute for Computational Engineering and Sciences, University of Texas at Austin, Austin, Texas 78712*

**Abstract.**

A recent experimental study reports on measuring the temporal duration and the spatial extent of failed attempts to cross an activation barrier (i.e., "loops") for a folding transition in a single molecule and for a Brownian particle trapped within a bistable potential. Within the model of diffusive dynamics, however, both of these quantities are, on the average, exactly zero because of the recrossings of the barrier region boundary. That is, an observer endowed with infinite spatial and temporal resolution would find that finite loops do not exist (or, more precisely, form a set of measure zero). Here we develop a description of the experiment that takes finite experimental resolution into account and show how the experimental uncertainty of localizing the point, in time and space, where the barrier is crossed leads to observable distributions of loop times and sizes. Although these distributions generally depend on the experimental resolution, this dependence, in certain cases, may amount to a simple resolution-dependent factor and thus the experiments do probe inherent properties of barrier crossing dynamics.

Recent single-molecule experiments have been able to observe, with great temporal and spatial resolution, how molecules cross an activation barrier *en route* between two metastable states such as the folded and unfolded states of a protein (see, e.g., refs. [1-4]). Until recently, such studies mainly focused on transition paths. A transition path is a segment of a molecular trajectory $x(t)$ that enters the barrier region $(a, b)$ through one of its boundaries (say $a$) and exits through the other $(b)$, as illustrated in Fig. 1. Properties of transition paths such as their temporal duration (i.e., the transition path time), average velocity and shape inform one about microscopic mechanisms of barrier crossing and offer an opportunity to test the applicability of various theories of barrier crossing; among those, the simple model of diffusive barrier crossing over a one-dimensional free energy barrier[5-8] is often found to provide a quantitative description of the process[9].

More recently, trajectories that enter the barrier region but do not necessarily traverse it have attracted attention. In particular, we studied[10], theoretically, the distributions of the exit time[11] for a system that starts somewhere within the barrier region. This time is a generalization of the transition path time, as it contains contributions both from transition paths and from nonreactive trajectories that enter and exit the barrier region through the same boundary. In the first (to our knowledge) experimental effort to probe barrier dynamics beyond transition paths[12], Lyons *et al*, measured the properties of nonproductive fluctuations, or "loops"[13-15], which enter and exit the barrier region through the same boundary (Fig. 1). In particular, they have measured two properties of loops. The first one is the distribution $p(x_{max}|a \to a)$ of the turning points $x_{max}$ for the loops that enter and exit the barrier region through the same boundary (here $a$). The second one is the distribution of the temporal duration of loops $p(t|a \to a)$.

For purely diffusive dynamics (with inertial effects ignored altogether, which is a good approximation at experimental timescales[16]), however, both of these distributions are pathological, and the exact expressions for them (see below for further details) are the delta functions:

$$p(x_{max}|a \to a) = 2\delta(x_{max} - a) \quad (1)$$

$$p(t|a \to a) = 2\delta(t) \quad (2)$$

Eqs. 1 and 2 simply state that a diffusive trajectory crossing the boundary $a$ will immediately recross back and thus will never penetrate the barrier region. In other words, the loops that start exactly at $a$ and have finite temporal duration/spatial extent form a set of measure zero. Experimental measurements, however, capture loops thanks to their limited spatial and temporal resolution: the exact moment when a boundary is crossed is unknown, and when the beginning of a loop is detected the trajectory is already a finite distance $\Delta$ away from the boundary (Fig. 1). As a result, the time it takes to return to the boundary and the distance the trajectory can travel into the barrier region are both finite. At first glance, then, one may conclude that the measured loop properties are experimental artifacts. The purpose of this note is (1) to explore the effect of experimental resolution on the measured loop properties and (2) to clarify why some of the loop properties measured in ref.[12] are essentially independent of the experimental resolution. This has already been recognized, at least qualitatively, by the authors of ref.[12], but here we propose a precise description of the apparent distributions $p(x_{max}|a \to a)$ and $p(t|a \to a)$ that take experimental uncertainties into account.

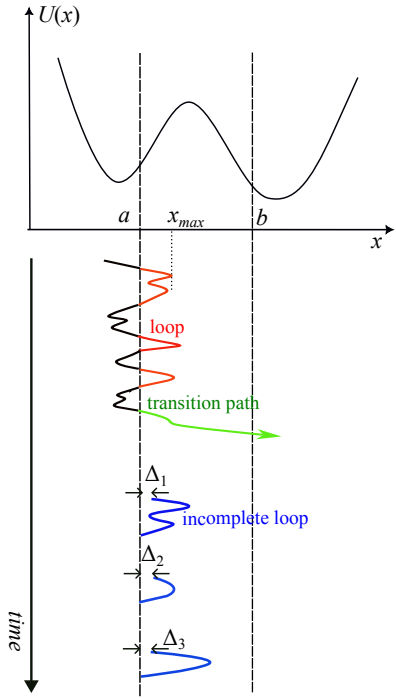

**Figure 1.** Top: here, we consider diffusive dynamics in a potential of mean force $U(x)$, with an interval $(a, b)$ identified as a barrier region. Bottom: A trajectory that enters the barrier region $(a, b)$ through its left boundary may ether proceed to exit through the right boundary forming a transition path (green) or exit through the left boundary resulting in a loop (red). Because of finite experimental resolution, a loop is detected a short time after it entered the barrier region, when it is located a distance $\Delta$ to the right of the boundary. Therefore, experiments measure properties of incomplete loops (blue) with their initial positions $x_{0,i} = a + \Delta_i$ (with $i$ enumerating trajectories) sampled from some distribution $\rho(x_0)$ determined by experimental details.

We consider the following model of an experiment performed with a limited spatio-temporal resolution. When an experimental trajectory $x(t)$ is being analyzed, each loop starts when it is observed, for the first time (say at $t = t_1$) to the right of the boundary $a$,

$x_0 \equiv x(t_1) > a$, having arrived from the region $x < a$. Because of finite spatial and temporal resolution, this staring point is slightly to the right of the boundary, $x_0 = a + \Delta > a$. As a result, one measures properties of "incomplete loops" starting to the right of the boundary (Fig. 1) rather than those of true loops. We will assume that the starting points of such incomplete loops are sampled from some resolution-dependent distribution $\rho(x_0)$, where the average distance of the starting point from the boundary,

$$\overline{\Delta} = \int_a^b \rho(x_0)(x_0 - a)dx_0, \quad (3)$$

characterizes the uncertainty in localizing the beginning of the crossing, or the fuzziness of the boundary as detected experimentally.

We now derive an expression for the probability density $p(x_{max}|x_0 \to a)$ of the turning points of incomplete loops by noting that $p(x_{max}|x_0 \to a)dx_{max}$ must be proportional to the fraction of trajectories that start at $x_0$ and reach $x_{max}$ but do not reach $x_{max} + dx_{max}$. This fraction is given by $\phi(x_0 \to x_{max}|a) - \phi(x_0 \to x_{max} + dx_{max}|a) = -\phi'(x_0 \to x_{max}|a)dx_{max}$, where $\phi(x_0 \to x|a)$ is the splitting probability, for a trajectory starting at $x_0 > a$, to reach $x$ before reaching $a$. Assuming coordinate-independent diffusivity, the latter is given by the known expression[17],

$$\phi(x_0 \to x|a) = \frac{\int_a^{x_0} dy\, e^{\beta U(y)}}{\int_a^x dy\, e^{\beta U(y)}} \quad (4)$$

and thus we have

$$\phi'(x_0 \to x|a) = -\frac{e^{\beta U(x)} \int_a^{x_0} dy\, e^{\beta U(y)}}{\left(\int_a^x dy\, e^{\beta U(y)}\right)^2} \quad (5)$$

Combining this with the obvious normalization requirement $\int_{x_0}^b p(x_{max}|x_0 \to a)dx_{max} = 1$, we obtain:

$$p(x_{max}|x_0 \to a) = F(x_{max})G(x_0), \; x_{max} > x_0, \quad (6)$$

with

$$F(x_{max}) = \frac{e^{\beta U(x_{max})}}{\left(\int_a^{x_{max}} dy e^{\beta U(y)}\right)^2} \quad (7)$$

and

$$G(x_0) = \frac{\left(\int_a^b dy e^{\beta U(y)}\right)\left(\int_a^{x_0} dy e^{\beta U(y)}\right)}{\int_{x_0}^b dy e^{\beta U(y)}} \quad (8)$$

Notice that if $x_0 = a$ then $G(x_0) = 0$ and so $p(x_{max}|a \to a) = 0$. This means that the system situated exactly at $a$ will never leave this boundary, and so the distribution of $x_{max}$ is formally the delta function, Eq. 1.

For small enough values of $x_0 - a$, corresponding to the starting points close to the boundary $a$, we can further approximate Eq. 8 by

$$G(x_0) \approx (x_0 - a)e^{\beta U(a)} \quad (9)$$

Since the exact location of the point $x_0$ relative to the boundary $a$ is below the experimental resolution, we should average Eq. 6 over the distribution $\rho(x_0)$ of the starting points. In doing so, it is only meaningful to consider points $x_{max}$ that are far enough from the fuzzy boundary set by the experimental resolution. Specifically, we assume $x_{max} - a \gg \bar{\Delta}$, where $\bar{\Delta}$, defined by Eq. 3, characterizes the distribution width; for such values of $x_{max}$ we are sure that the loop $x(t)$ turns around to the right of the point $x_0$ where it has started. The *apparent* distribution $\tilde{p}(x_{max}|a \to a)$ of the turning points (where the tilde indicates that we are referring to an apparent distribution) is then simply an average of Eq. 6 over the distribution of the starting points,

$$\tilde{p}(x_{max}|a \to a) = \langle G(x_0)\rangle_\rho F(x_{max}) \quad (10)$$

$$\langle G(x_0)\rangle_\rho = \int_a^{x_{max}} dx_0 \rho(x_0) G(x_0) \approx \int_a^b dx_0 \rho(x_0) G(x_0), \quad (11)$$

Using Eq. 9, we can further approximate the result as

$$\tilde{p}(x_{max}|a \to a) \approx \overline{\Delta} e^{\beta U(a)} F(x_{max}) = \overline{\Delta} \frac{e^{\beta U(x_{max}) + U(a)}}{\left(\int_a^{x_{max}} dy e^{\beta U(y)}\right)^2} \quad (12)$$

We emphasize that the factorization of Eq. 10 into $x_{max}$- and $x_0$-independent terms is only valid under the assumption that $x_{max} - a \gg \overline{\Delta}$. A key observation that ensues is that the apparent distribution of the turning points is always given by Eq. 7 multiplied by some numerical factor, regardless of the precise details of the measurement. This is precisely the observation made by Lyons et al[12], see Eq. 5 there.

We now turn to the distribution of loop times. Again, we consider the time duration of an incomplete loop $p(t|x_0 \to a)$ instead, assuming that the starting point is to the right of the boundary, $x_0 > a$. This is what is also known as the exit time[11] conditional upon reaching the boundary $a$ before reaching the boundary $b$. Properties of conditional exit time distributions have been studied in ref.[10], where formulas for its first and second moments were derived given the potential of mean force $U(x)$. Unfortunately, in this case the shape of this distribution (plotted in Fig. 4a of ref.[12]) depends, explicitly and nontrivially, on the starting point $x_0$ and/or on the distribution $\rho(x_0)$ of the starting points, and so the effect of the measurement cannot simply be reduced to a numerical factor, as in Eqs. 6, 10, and 12.

To illustrate this, let us consider very short loops (corresponding to the case where $x_0$ and $x_{max}$ are sufficiently close to the boundary $a$). As such short loops cannot travel very far from the boundary, one can neglect the effect of the potential $U(x)$ and use the approximation[10] where $U(x)$ is constant. Moreover, one can ignore the existence of the right boundary $b$, which is unlikely to be reached during a short loop time. This results in a well known formula[11]

$$p(t|x_0 \to a) \approx \frac{(x_0-a)\exp-\frac{(x_0-a)^2}{4Dt}}{2\sqrt{\pi D t^3}} \quad (13)$$

Unlike Eq. 6, this distribution depends on the location of the initial point $x_0$ in a nontrivial way. This suggests that the properties of the measured distribution of the temporal loop duration cannot be understood without explicit consideration of the experimental uncertainties and the precise manner in which the trajectories are analyzed.

An interesting feature of Eq. 13 is that this is a broad distribution, with a power-law tail. This is in agreement with the experimentally measured distribution, see ref.[12], Fig. 4a, and in contrast with the narrow distributions expected for the transition path times in the case of diffusive dynamics[18]. The first moment of the distribution of Eq. 13 (as well as its higher moments) diverges, but this divergence is removed when the second boundary, $x = b$, is taken into account[10], as this boundary limits the time the system can spend in the barrier region.

The sensitivity of the loop time distribution to experimental uncertainties makes it difficult to interpret it as a fundamental property of the observed dynamics. One alternative is to consider the distribution of the loop time conditional upon having $x_{max}$ as the loop's turning point,

$$p[t|x_0 \to a, \max_{0<t'<t} x(t') = x_{max}] \equiv p(t|x_0 \to x_{max} \to a) \quad (14)$$

We note in passing that this conditional distribution, together with the distribution of the turning points, $p(x_{max}|x_0 \to a)$, contains, in principle, all information about the

unconditional distribution $p(t|x_0 \to a)$, as the latter can (in principle) be obtained from $p(t|x_0 \to x_{max} \to a)$ by averaging over $x_{max}$,

$$p(t|x_0 \to a) = \int_{x_0}^{b} dx_{max}\ p(x_{max}|x_0 \to a)p(t|x_0 \to x_{max} \to a), \quad (15)$$

Unlike the unconditional distribution $p(t|x_0 \to a)$, the conditional distribution $p(t|x_0 \to x_{max} \to a)$ is only weakly dependent on the starting point $x_0$, and the limit

$$\lim_{x_0 \to a} p(t|x_0 \to x_{max} \to a) = p(t|a \to x_{max} \to a) \quad (16)$$

is well behaved for $x_{max} \gg a + \overline{\Delta}$, which is the case assumed here. For this reason, for a sufficiently small values of $\overline{\Delta}$ it is not necessary to differentiate between the distribution $p(t|x_0 \to x_{max} \to a)$ for incomplete loops and $p(t|a \to x_{max} \to a)$ for complete loops, and it can be assumed that the experimental measurement directly yields $p(t|a \to x_{max} \to a)$.

Each $a \to x_{max} \to a$ loop further consists of a transition path from $a$ to $x_{max}$ and a transition path from $x_{max}$ to $a$. Therefore, measuring the time spent by the system on such a loop amounts to measuring the sum of two statistically independent (if obeying Markovian dynamics) transition path times (with the boundaries $a$ and $x_{max}$) and is thus analogous to earlier transition path measurements. The distribution of this time is a convolution of two transition path time distributions:

$$p(t|a \to x_{max} \to a) = \int_0^t d\tau\, p(\tau|a \to x_{max})p(t-\tau|x_{max} \to a) = \int_0^t d\tau p(\tau|a \to x_{max})p(t-\tau|a \to x_{max}), \quad (17)$$

where we recognized that the time-reversal symmetry of transition paths[19-21] leads to identity of distributions $p(t|a \to x_{max})$ and $p(t|x_{max} \to a)$ for transition paths from $a$ to $x_{max}$ and from $x_{max}$ to $a$.

Recent work highlighted the shapes of the distributions of barrier crossing times as possible signatures of microscopic dynamics[10, 18, 22-25]. In particular, the shape of the transition path time distribution is always narrow in the case of diffusive dynamics, such that its standard deviation is always smaller than its mean. This property is often quantified by the value of the distribution's coefficient of variation $C$, which is less than 1 for diffusive activate rate processes:

$$C = \frac{\sqrt{\langle t^2 \rangle - \langle t \rangle^2}}{\langle t \rangle} < 1, (18)$$

and thus violation of this inequality indicates breakdown of the Kramers-type picture of diffusive barrier crossing[3, 26, 27] resulting, e.g., from multiple distinct transition pathways. Here, for any distribution $p(t)$, the moment $\langle t^n \rangle$ is defined by

$$\langle t^n \rangle = \int_0^\infty dt\, t^n p(t). (19)$$

Other times related to barrier dynamics, such as exit times, however, can have broad distributions even in the case of diffusive dynamics[10]. In this light, it is instructive to compare the shape of the unconditional distribution $p(t|x_0 \to a)$ with that of the conditional loop time distribution $p(t|a \to x_{max} \to a)$. As mentioned above and discussed in ref.[10], the former is broad and can violate Eq. 18. In contrast, the latter is always narrow, with a coefficient of variation $C$ satisfying Eq. 18. Let us outline the proof that $C < 1$ in this case.

First we note that Eq. 18 has been proven earlier for transition path time distributions[18, 22, 23] (assuming diffusive dynamics). Second, it is easy to prove that, given that $p(\tau|a \to x_{max})$ satisfies Eq. (18), the convolution of this distribution with itself, Eq. 17, also satisfies Eq. 18. In fact, if the coefficient of variation of the distribution $p(t|a \to x_{max})$ is $C(<1)$, then the coefficient of variation of the $p(t|a \rightarrow x_{max} \rightarrow a\ )$, given by the convolution of Eq. 17, is $C' = \frac{C}{\sqrt{2}} < 1$. This can be shown, for example, by expressing the moments of each distribution as $\langle t \rangle = -\hat{p}'(0), \langle t^2 \rangle = \hat{p}''(0)$, where $\hat{p}(s) = \int_0^\infty e^{-st} p(t) dt$ is the Laplace transform of $p(t)$, and by writing the Laplace transform of $p(t|a \rightarrow x_{max} \rightarrow a\ )$, which is a convolution, as the product of the two identical Laplace transforms, i.e., $\hat{p}^2(s|a \rightarrow x_{max}\ )$.

To summarize, unlike transition paths, whose properties are relatively insensitive to the fuzziness of the experimental barrier boundaries, failed barrier crossing attempts, or loops, have pathological properties within the model of diffusive dynamics: a loop that starts at a boundary of a barrier region will never leave this boundary. The experimental uncertainty in locating the precise crossing of the boundary results in loops of finite spatial size and of finite temporal duration, but this raises the question of whether such loops are experimental artifacts or whether they report on inherent properties of the dynamics within the barrier region. Here we have shown that certain properties of loops thus measured, such as the shape of the distribution of the turning points $x_{max}$, are independent of or weakly dependent on the measurement accuracy. This observation offers a precise theoretical foundation to experimental studies of failed barrier crossing attempts. In particular, two experiments performed on the same system but with instruments of different resolution will measure the same distribution of the turning points (proportional to the function $F(x_{max})$, see Eqs. 6 and 10) to within a numerical factor. This function, then, can be thought of as an inherent property of barrier crossing dynamics. In contrast, the experimentally observable distribution of (unconditional) loop times does not have such a simple interpretation, as it depends on the experimental resolution in a nontrivial way. A related function, the distribution of loop times conditional on the position $x_{max}$ of the turning point, is well behaved but it less interesting (at least in the case of Markovian dynamics), as it provides information that is also easily obtainable by considering transition paths over a modified barrier region $(a, x_m)$.

We note that finite experimental resolution will also lead to additional errors in measuring the duration of failed barrier crossing attempts: for example, a trajectory sampled at finite time intervals may cross and recross a boundary thus terminating a loop, yet this crossing may be unobserved, resulting in a larger apparent loop time. This kind of systematic errors is not considered here, but it has been the subject of recent work[28-30].

## Acknowledgements

AMB was supported by the Intramural Research Program of the NIH, *Enice Kennedy Shriver* National Institute of Child Health and Human Development. DEM was supported by the Robert A. Welch Foundation (Grant No. F- 1514) and the National Science Foundation (Grant No. CHE 1955552).

## References

1. H. S. Chung and W. A. Eaton, Curr Opin Struct Biol **48**, 30-39 (2018).
2. N. Q. Hoffer and M. T. Woodside, Curr Opin Chem Biol **53**, 68-74 (2019).
3. D. E. Makarov, J Phys Chem B **125** (10), 2467-2476 (2021).
4. H. S. Chung and I. V. Gopich, Phys Chem Chem Phys **16** (35), 18644-18657 (2014).
5. H. A. Kramers, Physica **7**, 284-304 (1940).
6. A. Szabo, K. Schulten and Z. Schulten, J Chem Phys **72**, 4350 (1980).

7. N. D. Socci, J. N. Onuchic and P. G. Wolynes, J. Chem. Phys. **104**, 5860-5868 (1996).
8. D. Klimov and D. Thirumalai, Phys. Rev. Lett. **79**, 317 (1997).
9. K. Neupane, A. P. Manuel and M. Woodside, Nature Physics **12**, 700-703 (2016).
10. A. M. Berezhkovskii and D. E. Makarov, Biophysical Reports **1**, 100029 (2021).
11. S. Redner, *A Guide to First Passage Times*. (Cambridge University Press, 2001).
12. A. Lyons, A. Devi, N. Q. Hoffer and M. T. Woodside, Physical Review X **14** (1), 011017 (2024).
13. A. M. Berezhkovskii, L. Dagdug and S. M. Bezrukov, J Chem Phys **147** (13), 134104 (2017).
14. A. M. Berezhkovskii, L. Dagdug and S. M. Bezrukov, J Phys Chem B **121** (21), 5455-5460 (2017).
15. A. M. Berezhkovskii, L. Dagdug and S. M. Bezrukov, J Phys Chem B **123** (17), 3786-3796 (2019).
16. A. M. Berezhkovskii and D. E. Makarov, J Chem Phys **148** (20), 201102 (2018).
17. C. W. Gardiner, *Handbook of Stochastic Methods for Physics, Chemistry and the Natural Sciences*. (Springer-Verlag, Berlin, 1983).
18. R. Satija, A. M. Berezhkovskii and D. E. Makarov, Proc Natl Acad Sci U S A **117** (44), 27116-27123 (2020).
19. A. M. Berezhkovskii, G. Hummer and S. M. Bezrukov, Phys Rev Lett **97** (2), 020601 (2006).
20. S. Chaudhury and D. E. Makarov, J. Chem. Phys. **133**, 034118 (2010).
21. D. E. Makarov, *Single Molecule Science: Physical Principles and Models*. (CRC Press, Taylor & Francis Group, Boca Raton, 2015).
22. A. M. Berezhkovskii, S. M. Bezrukov and D. E. Makarov, J Chem Phys **154** (11), 111101 (2021).
23. D. Hartich and A. Godec, Phys. Rev. X **11**, 041047 (2021).
24. R. Dutta and E. Pollak, Phys Chem Chem Phys **23** (41), 23787-23795 (2021).
25. R. Dutta and E. Pollak, Phys Chem Chem Phys **24** (41), 25373-25382 (2022).
26. R. Satija, A. Das, S. Muhle, J. Enderlein and D. E. Makarov, J Phys Chem B **124** (17), 3482-3493 (2020).
27. A. Godec and D. E. Makarov, J Phys Chem Lett **14** (1), 49-56 (2023).
28. D. E. Makarov, A. M. Berezhkovskii, G. Haran and E. Pollak, J. Phys. Chem. B **126** (40), 7966–7974 (2022).
29. K. Song, D. E. Makarov and E. Vouga, The Journal of Chemical Physics **158** (11), 111101 (2023).
30. A. Kumar, Y. Scher, S. Reuveni and M. S. Santhanam, Physical Review Research **5** (3), L032043 (2023).